\documentclass[letter,12pt]{article}
\usepackage[T1]{fontenc}
\usepackage[english,spanish]{babel}
\usepackage{amsmath}
\usepackage{amssymb,amsfonts,textcomp}
\usepackage{array}
\usepackage{hhline}
\usepackage[pdftex]{graphicx}
\usepackage{mathrsfs}

\usepackage[latin1]{inputenc} 
\usepackage{amsfonts,amsthm}

\theoremstyle{definition}

\numberwithin{equation}{section}
\newtheorem{teorem*}[thrm]{Theorem}
\usepackage{color}

\evensidemargin = \oddsidemargin
\newcommand{\nada}[1]{}

\begin{document}

\title{Estimación de población contagiada por Covid-19 usando regresión Logística generalizada y\\ heurísticas de optimización}

\author{ Villalobos Arias, Mario Alberto\thanks{Universidad de Costa Rica, CIMPA y Escuela de Matemática, San José, Costa Rica, mario.villalobos@ucr.ac.cr\newline Instituto tecnologico de Costa Rica, Escuela de Matemática, Cartago, Costa Rica, marvillalobos@itcr.ac.cr} 
}

\bigskip

\maketitle

\begin{abstract}
En este trabajos se presenta una propuesta para la estimación de la poblaciones usando ajuste de curvas del tipo logística.
Este tipo de curvas se utilizan para el estudio de crecimiento de poblaciones, en este casos población de personas infectadas por el virus Covid-19; y también se puede utilizar para aproximar la curva de supervivencia que se utiliza en estudios actuariales y otras similares.
\end{abstract}


\section{Introducción} 
\begin{figure}[ht]
\centerline{\includegraphics[width=9cm,height=6cm]{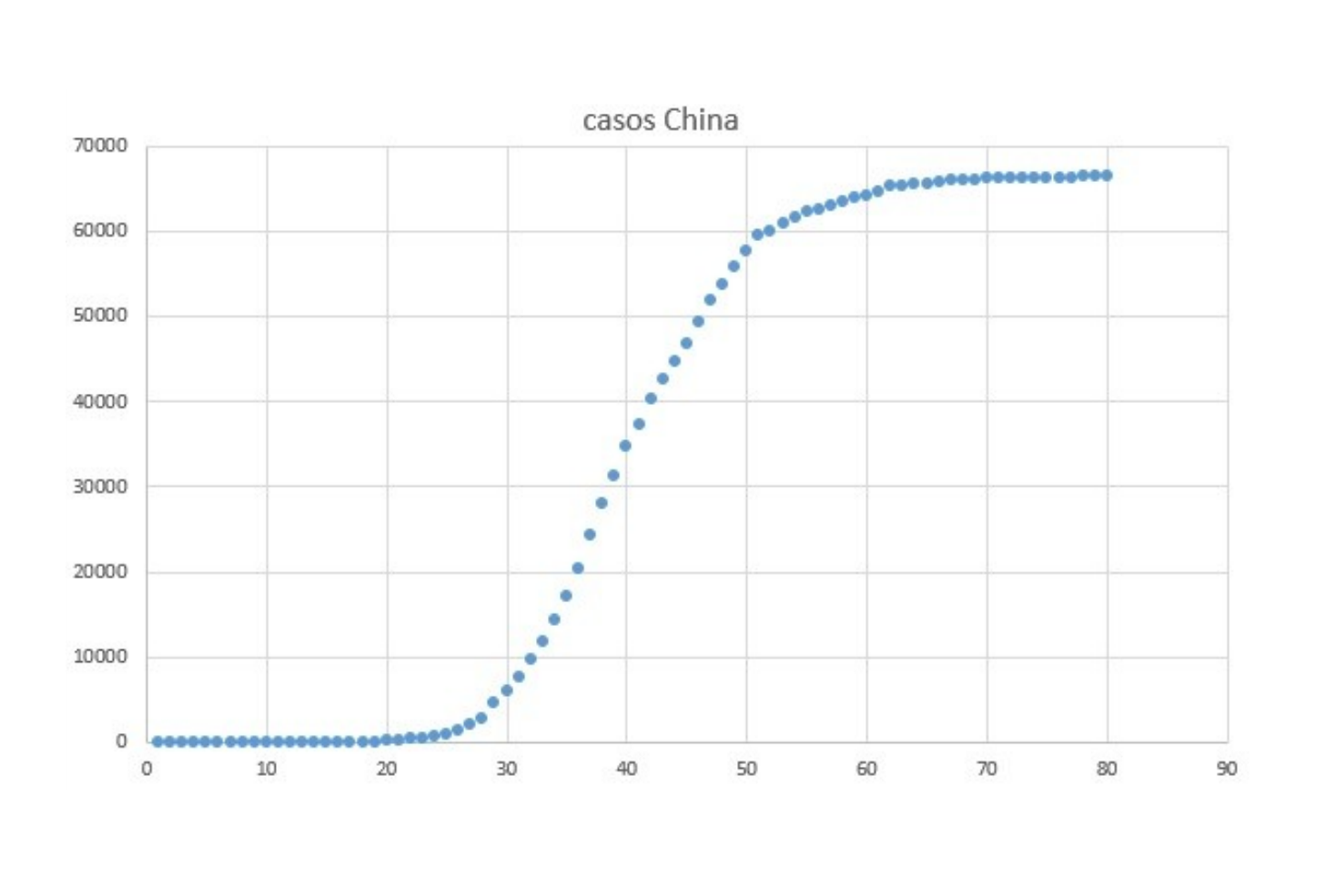}}
\caption{casos totales de contagiados en China}
\label{datosChina}
\end{figure}

Las curvas de crecimiento de poblaciones siguen el conocido comportamiento logístico como se muestra en la figura \ref{datosChina}.

Un modelo que se utiliza para el ajuste de curvas de poblaciones es el Logístico, que se utiliza la siguiente ecuación
\begin{equation}
P(t) \ = \ \frac{1}{1+e^{-at+b}}
\label{logistica}
\end{equation}

Como se ve este modelo tiene varios problemas entre ellos que los datos están en $[0,1]$, y no es flexible, la ventaja es que este modelo es que se puede obtener una aproximación de la solución óptima al transformar los datos y utilizar regresión lineal.

\begin{figure}[ht]
\centerline{\includegraphics[width=9cm,height=6cm]{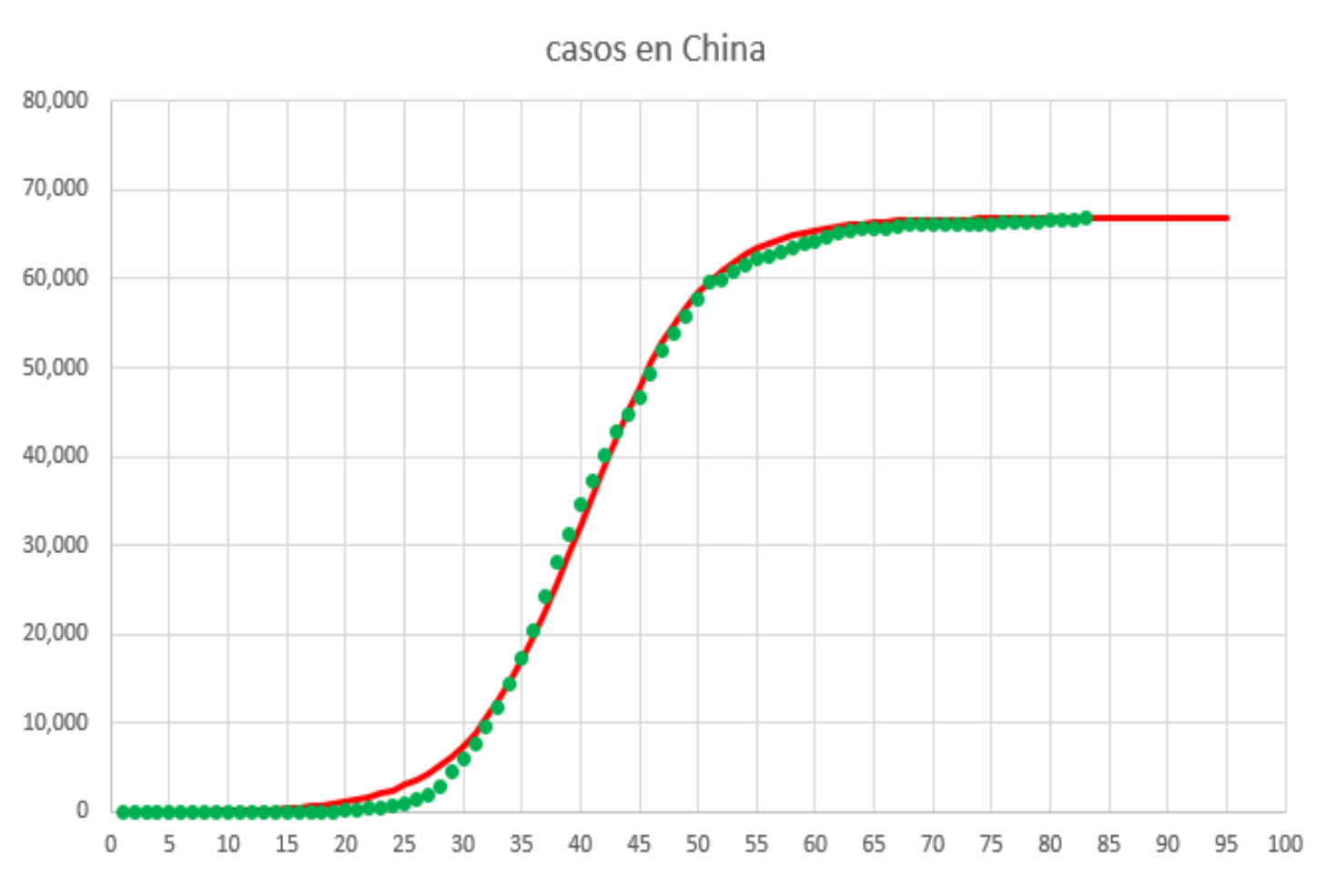}}
\caption{Ajuste con regresión logística totales de contagiados en China}
\label{logChina}
\end{figure}

Como se ve en el gráfico \ref{logChina} al aplicar regresión logística 
y al dividir todos los datos entre el máximo valor (66818) se obtiene una curva que ajusta bien, con un $R^2=0.99955$, que es muy bueno,  desde el punto vista estadístico, pero como se ve en la figura hay muchos valores que no ajustan muy bien en las curvas y no es bueno para la predicción, como veremos más adelante.

\section{El modelo modelo SIR}

La versión clásica para el estudio de epidemias es el modelo modelo SIR en el cual  se divide a la población en tres grupos: los susceptibles, los infectados y los recuperados (SIR), esto en el caso más simple y que la población va cambiando de susceptible a infectados y luego recuperados.

Se supone que el decrecimiento  es proporcional a la cantidad de infectados multiplicado por  la misma cantidad de susceptibles.
el cambio en los recuperados es igual a cierto porcentaje de los infectados y finalmente 
la cantidad de infectados va a cambiar aumentando por los susceptibles y luego le quitamos la cantidad de los infectados que pasa recuperados. De esta manera se obtienen las siguientes ecuaciones

$$
{\displaystyle {\begin{aligned}&{\frac {dS}{dt}}=-{\frac {\beta IS}{N}},\\[6pt]&{\frac {dI}{dt}}={\frac {\beta IS}{N}}-\gamma I,\\[6pt]&{\frac {dR}{dt}}=\gamma I,\end{aligned}}}
$$
con
$$
{\displaystyle {\frac {dS}{dt}}+{\frac {dI}{dt}}+{\frac {dR}{dt}}=0,}
$$
que nos da 
$$
{\displaystyle S(t)+I(t)+R(t)={\text{constant}}=N,}
$$

El problema con el modelo SIR es que como vemos se necesitan los tres parámetros que se tienen en las ecuaciones diferenciales pero con pocos datos o digas en pocos días es muy difícil determinar esos  parámetros

\section{Regresión Logística generalizada}

Por lo anterior se proponen la utilización de un modelo un poco más complejo a la regresión logística y que sea más fácil de determinar que el SIR

Una primera versión es:
\begin{equation}
P(t) \ = \ \frac{M}{1+e^{-at+b}}
\label{logGenM}
\end{equation}
que agrega un parámetro más, por determinar, que es la población límite $M$, ya con esta modificación no se puede resolver mediante transformación y regresión lineal. Para resolver este problema se debe utilizar técnicas no lineales de optimización para resolverlo, además tiene el problema de rigidez, esto es que no se ajusta suficientemente a ciertas curvas, este modelo se puede utilizar para predicción, pero desde el punto de ajuste no es tan preciso como veremos más adelante.
 
Para flexibilizar la curva se agrega un parámetro extra $\alpha$ de la siguiente forma 
\begin{equation}
P(t) \ = \ \frac{M}{\big(1+e^{-at+b}\big)^{\alpha}}.
\label{logGen}
\end{equation}

Este parámetros $\alpha$ agrega flexibilidad en el ajuste, recuérdese las gráficas de $y=x^{1/3}$, $y=x^{1/2}$, $y=x$,$y=x^2$,$y=x^3$.

Para esta función el punto de inflexión es cuando 

$$
P''(t) \ = \ a^2 c e^{a t + b} (e^{a t + b} + 1)^{-c - 2} (c e^{a t + b} - 1) \ = \ 0
$$
 por lo que se tiene que el punto de inflexión se obtiene para 
$$
t= -\frac{\ln (\alpha)+b}a
$$

Este es el mismo modelo conocido como la curva de Richards que se utiliza para modelar crecimientos de poblaciones. 
\begin{equation}
Y(t)\ = \ A+{K-A \over (C+Qe^{{-Bt}})^{{1/\nu }}}
\end{equation}
Note que con algunos cálculos y  $A = 0$ se da la igualdad con la logística generalizada

El caso:
$$
Y(t)={K \over (1+Qe^{{-\alpha \nu (t-t_{0})}})^{{1/\nu }}}
$$
que es solución de la ecuación diferencial:
$$
Y^{{\prime }}(t)=\alpha \left(1-\left({\frac  {Y}{K}}\right)^{{\nu }}\right)Y
$$
\subsection{Función de Gompertz}
debe su nombre a Benjamin Gompertz, el primero en trabajar en este tipo de función es un caso particular de la de Richards, y tiene por ecuación la siguiente:  
$$
{\displaystyle G(t)=a\mathrm {e} ^{-b\mathrm {e} ^{-ct}},}
$$
Además su segunda derivada es:
$$
G''(t) \ = \ b c^2 e^{b e^{c x} + c x} (1 + b e^{c x}) \ = \ 0
$$
que nos da el punto de inflexión se alcanza en $t= -log(-b))/c$
lo que en el caso de epidemias nos dice en que punto el crecimiento de los casos diario va a iniciar a descender.

Esta es una función más simple pues solo tiene 3 parámetros en vez de la LG, que tiene 4, y por ende va a tener menos óptimos locales.

\subsection{Otras versiones}
Otras versiones más complejas son 
\begin{equation}
P(t) \ = \ \frac{M + ct}{\big(1+e^{-at+b}\big)^{\alpha}}.
\label{logGenrecta}
\end{equation}

Al momento de escribir este trabajo, los datos de covid-19, de Korea del sur se tiene que ajustar con una curva como esta.

Y la siguiente que se utilizó para ajustar las curvas de supervivencia 
\begin{equation}
P(t) \ = \ \frac{M}{\big(1+e^{-at^2+bt+c}\big)^{\alpha}},
\label{logGenCuad}
\end{equation}

\section{Hipótesis y propuesta}

La propuesta de este trabajo es, primero  utilizar la curva logística generalizada o la curva de Gompertz para hacer un ajuste de los datos en los que se se tengan la curva casi completa, por ejemplo datos de China y de Corea del Sur,(30 de marzo)

Por otro lado cuando se tengan todos los datos existe una curva del tipo de regresión logística generalizada o de Gompertz que ajusta los datos.

Por lo que la hipótesis aquí es: 
Si yo tengo la parte bajo de la  Curva, es decir los primeros valores de la curva, yo puedo obtener los parámetros de la curva, y obtener la curva completa. Y con esto se puede utilizar para predecir el crecimiento de la población, en este caso, por ejemplo, el número total de casos por el covid-19 en un país o región y cuando se llega al punto de inflexión, es decir, momento en que el número de casos diarios empieza a descender.

Más específicamente si se tiene los datos de unos 20 a 30 ó 35 días la pregunta es:
 
¿se puede determinar los parámetros de la curva que ajusta los datos completos?

Si esto se logrará como vemos tendríamos una forma de predecir el comportamiento del crecimiento de poblaciones con sólo tener unos pocos días, es decir podemos  determinar los parámetros de la curva con pocos días, que es muy difícil con el modelo SIR determinar los parámetros de las ecuaciones diferenciales.

\section{Ajuste de curvas}

Para este trabajo se están utilizando los datos provistos por el European Centre for Disease Prevention and Control en la página para bajar los datos diaios ver \cite{datos}.

Lo primero que se presenta es verificar, como se sabe, que la curva LG ajusta muy bien los datos de covid-19. Para el caso de China y Korea del Sur, que son los que se tiene la curva casi completa.

Utilizando un algoritmo de optimización no lineal se tienen los siguientes resultados.

\subsection{China: datos original}
\label{ChinaO}
\begin{figure}[ht]
\centerline{\includegraphics[width=9cm,height=6cm]{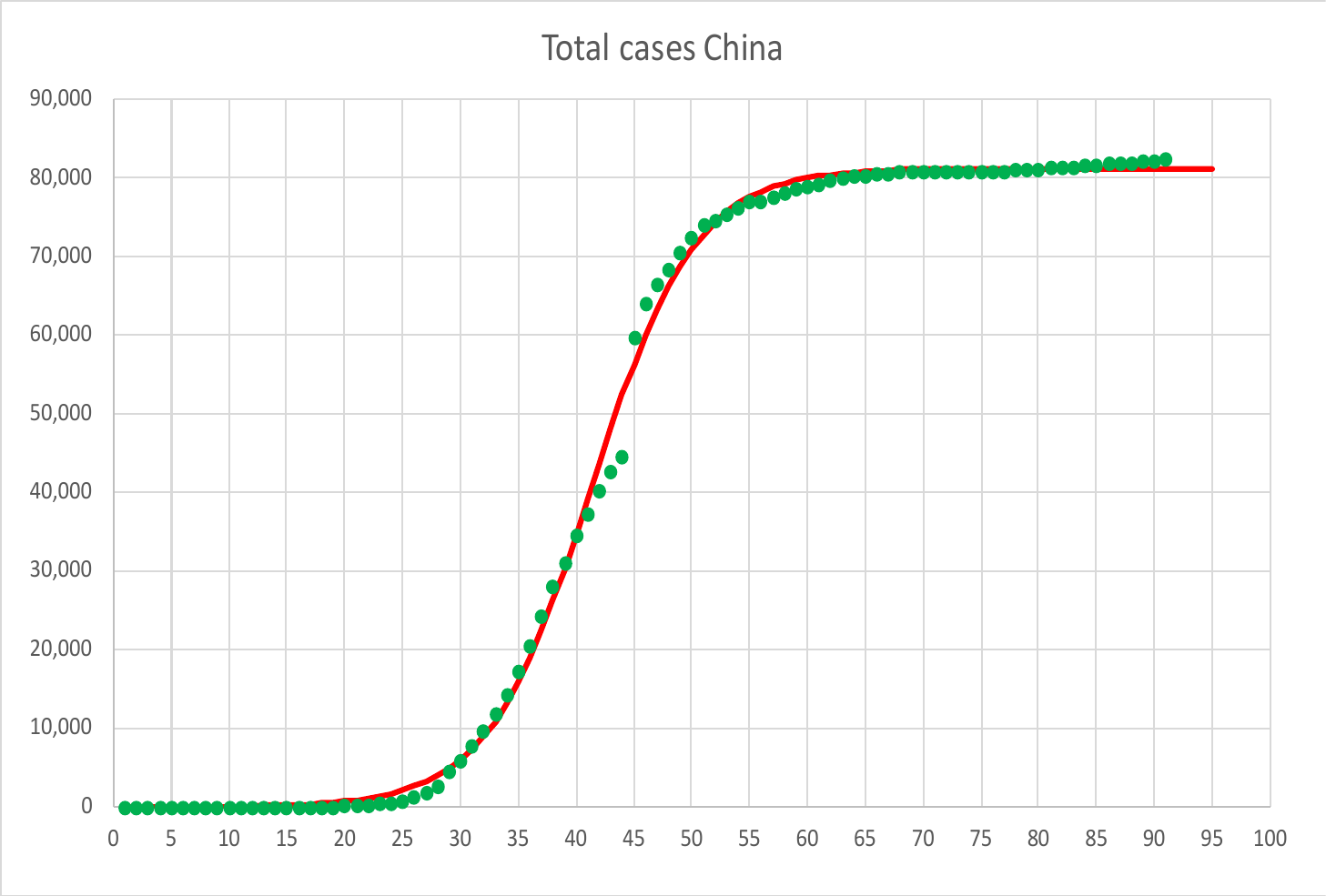}}
\caption{Ajuste con regresión logística totales de contagiados en China}
\label{ChinaOri}
\end{figure}

Para los datos de China se obtiene los siguientes datos y el gráfico en la figura \ref{ChinaOri}.

$$
\begin{array}{ccccc}
fecha &  	M &  	a &  	b &  		\alpha \\\hline 	
31/03/2020 &  	81149 &  	-0.2348563&  	9.89996092 &  	0.8809852 \\\hline 	
\end{array}
$$
Con $R^2=	0.998878.$
Como se observa se obtiene una muy buena aproximación, pero como sabemos y se ve en el la figura \ref{ChinaOri}, hay un salto en esos datos, por lo que se decidió corregir ese salto poniendo
 del dato del día 13/02/2020 igual al día anterior y el del día
14/02/2020 igual al día siguiente.

\begin{figure}[ht]
\centerline{\includegraphics[width=9cm,height=6cm]{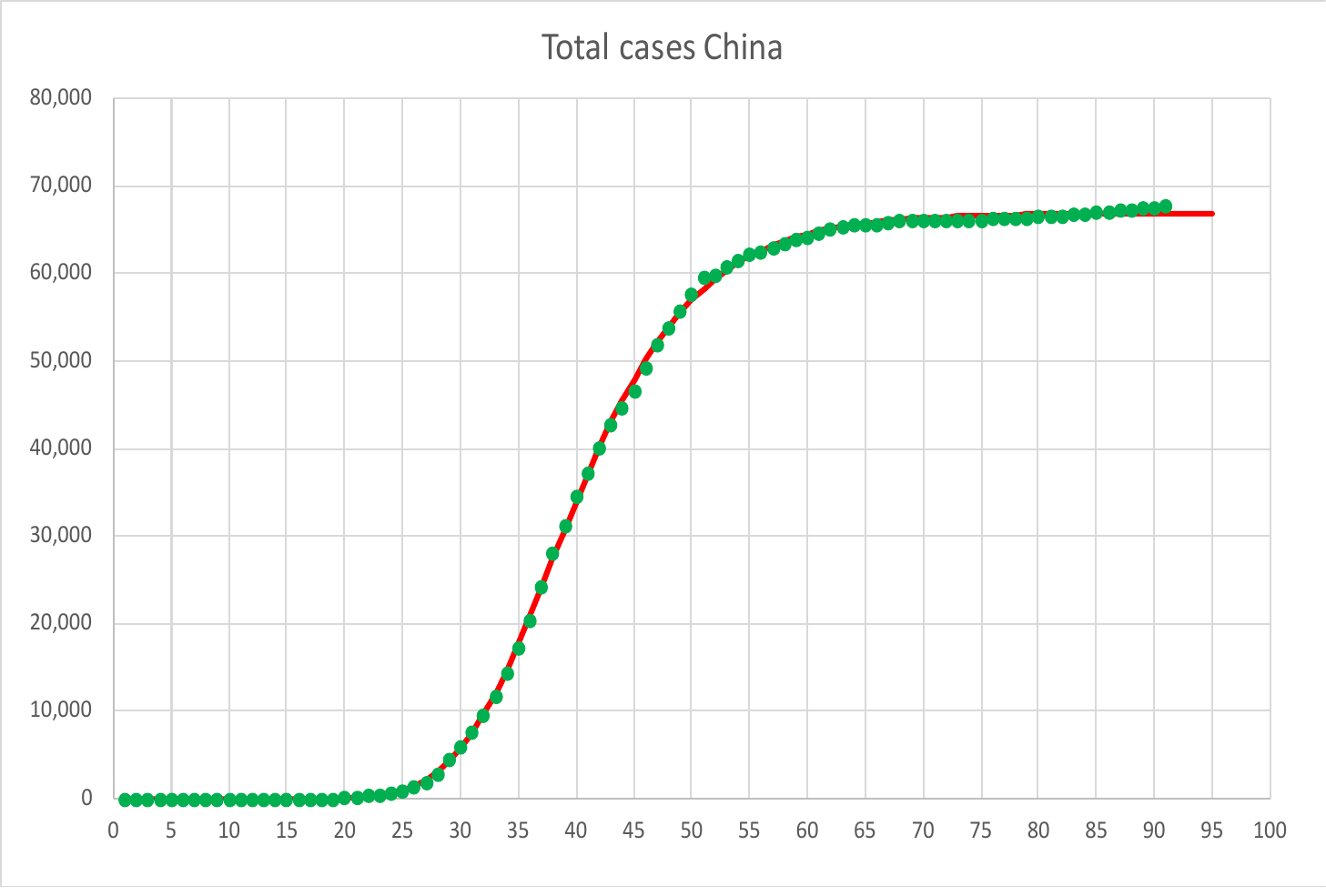}}
\caption{Ajuste con regresión logística generalizada en China {\bf datos corregidos}}
\label{ChinaCorr}
\end{figure}

Con estas correcciones se obtiene los siguientes datos y el gráfico de la figura \ref{ChinaCorr}

$$
\begin{array}{ccccc}
fecha &  	M &  	a &  	b &  		\alpha \\\hline 
31/03/2020	 &  66871 &  	-0.15032863 &  	4.21103499 &  		4.436336 \\\hline
\end{array}
$$
En este caso se obtiene un $R^2= 0.999885 $, que es mejor al de los datos sin corrección.

Observe que en los últimos datos se ve que hay una tendencia lineal, esto se puede mejorar comos se ve en el caso de Corea del Sur, como se ve en la siguiente subsección.

\subsection{Corea del Sur: Datos Originales}
\label{CoreaO}
\begin{figure}[ht]
\centerline{\includegraphics[width=9cm,height=6cm]{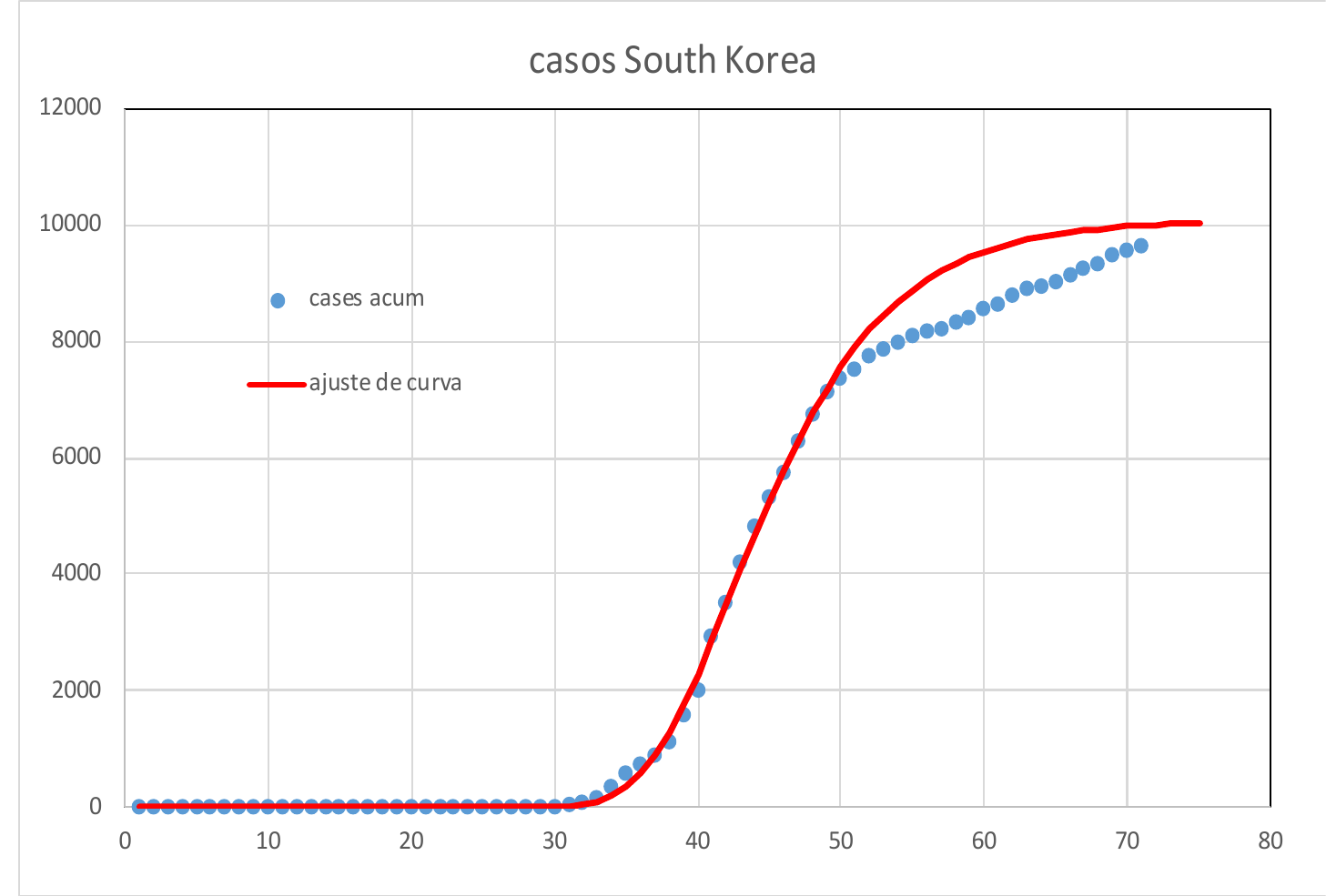}}
\caption{Ajuste con regresión logística generalizada  Corea del Sur}
\label{CoreaSur}
\end{figure}

Al aplicar el método a los datos de Corea del Sur se obtiene los datos siguientes y el gráfico de la figura \ref{CoreaSur}.
$$
\begin{array}{ccccc}
fecha &  	M &  	a &  	b &  		\alpha \\\hline 
31/03/2020	 &  10079 &  	-0.16416438 &  	0.18353146 &  		874.692503   \\\hline
\end{array}
$$
Y se obtiene un $R^2=	0.99875$, que no es muy bueno como se ve en la figura \ref{CoreaSur} que el ajuste no es muy bueno.

Para mejorar este ajuste se propone utilizar la curva de ajuste  (\ref{logGenrecta}), a saber:

$$
P(t) \ = \ \frac{M + ct}{\big(1+e^{-at+b}\big)^{\alpha}}.
$$
Con esta curva se logra mejorar el ajuste al obtener los siguientes datos:
$$
\begin{array}{cccccc}
fecha &  	c &  	M &  	a &  	b &  		\alpha \\\hline 
31/03/2020	 &  92.63407116 &  	3046 &  	-0.36922771 &  	15.5617475 &  		0.9691940884 \\\hline
\end{array}
$$

\begin{figure}[ht]
\centerline{\includegraphics[width=9cm,height=6cm]{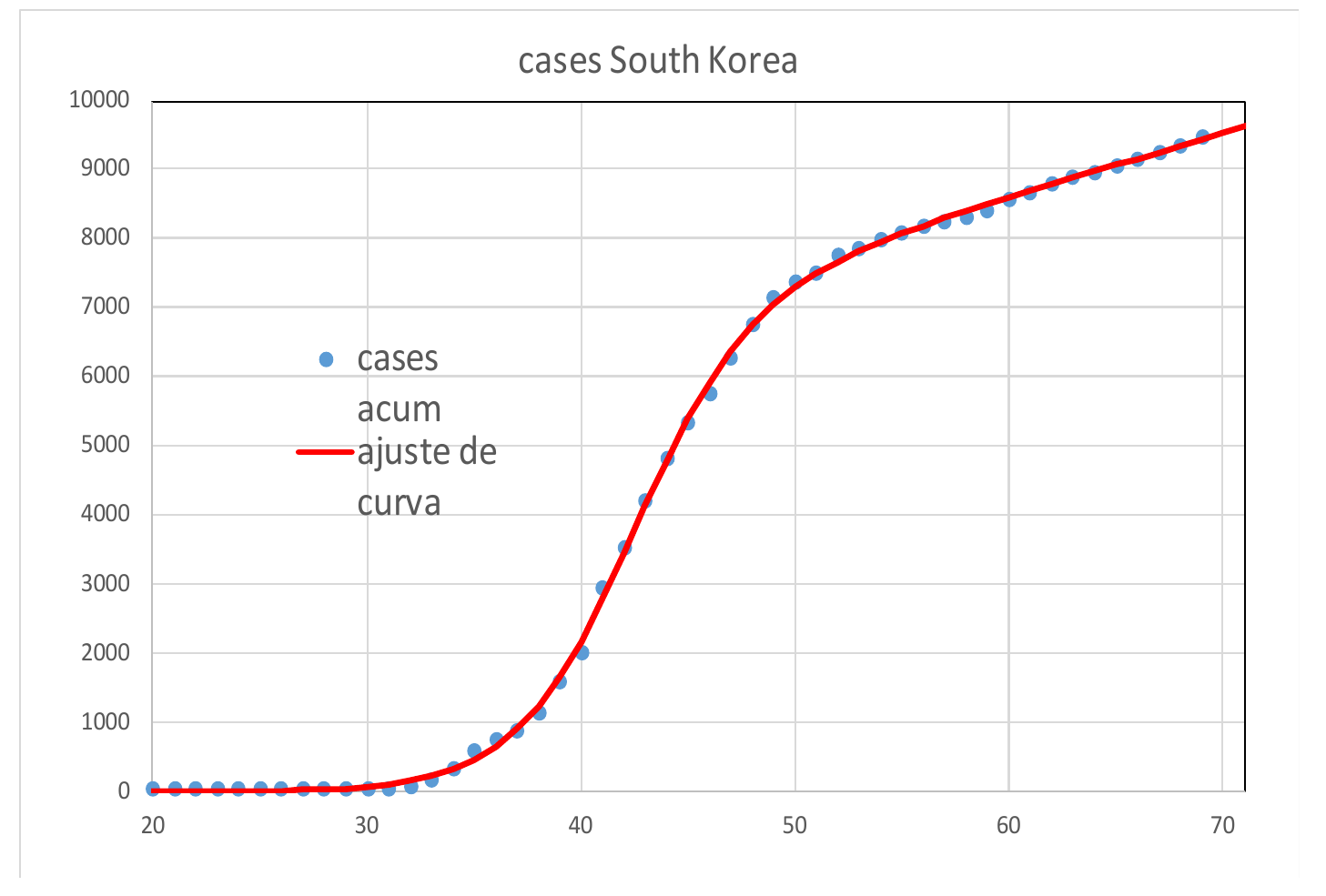}}
\caption{Ajuste con regresión logística generalizada de contagiados en Corea del Sur {\bf con tendencia final lineal}.}
\label{CoreaSurRecta}
\end{figure}

Y ahora el $R^2=	0.99988	$ que es mejor al resultado anterior y como se ve en el gráfico	de la figura \ref{CoreaSurRecta}.
	
Como se ve el ajuste con este tipo de curvas logísticas generalizadas es muy bueno obteniendo $R^2$ superiores a 0.9999.

\section{Pruebas de Predicción}
\label{PbaPred}

Como se vio en la sección anterior las curvas logísticas aproximan muy bien, como se ha demostrado en trabajos anteriores  (por ejemplo \cite{Fekedulegn, Pella}).

En esta sección vamos a tratar de medir el porcentaje de error relativo que se comete al utilizar las curvas propuestas, se utilizará una curva de prueba y los datos de China y Corea del Sur

\subsection*{Curva de prueba}

Para tratar de validar  la hipótesis vamos a generar una curva con valores 
$$
\begin{array}{cccc}
 	M &  	a &  	b &  		\alpha \\  \hline
  2000 &  	-0.05 &  	-2&  		10 \\\hline  
\end{array}
$$

y al utilizar el algoritmo de optimización se obtiene lo siguientes cotas superiores para el error relativo 
al usar la función LG, con los días indicados:

\begin{center}
{\sc Porcentaje de error relativo en curva de prueba\\ usando funcion Logística Generalizada}

\begin{tabular}{llr}
número de días &  máx error rel.\\\hline
25 días	&		10.451\% \\	
30 días	&		4.801\%	\\
35 dias	&	1.825\%\\
40 días		&	1.188\%	\\\hline
\end{tabular}
\end{center}

Al utilizar la función de Gompertz se mejora notablemente la predicción, como se había mencionado por tener menos parámetros, 
al utilizar el ejemplo con 
$$
\begin{array}{cccc}
 	a &  	b &  		c \\ \hline 
  2000 &   	-2&  		10   
\end{array}
$$
Al ejecutar el método se  obtiene los siguientes porcentajes de error relativo

\begin{center}
{\sc Porcentaje de error relativo en curva de prueba\\ usando Gompertz}

\begin{tabular}{cc}
número de días &  máx error rel.\\\hline
25 días	&		1.6697\% \\	
30 días	&		0.7492\%	\\
35 dias	&	0.8928\%\\
40 días		&	0.3234\%	\\\hline
\end{tabular}
\end{center}

Como se ve con la función de Gompertz, al parecer, se puede garantizar mejor predicciones.

\subsection*{Datos de China}

Veamos ahora el caso de China, en el que se tiene la curva casi completa.

Como vemos al utilizar los días del 16 al  45 (20 días), observe que los primeros 16 días casi no hubo cambios en los datos por lo que no se utilizaran esos datos, con estos datos con LG se obtiene:

\subsection*{Gompertz}

Al utilizar los días del 16 al  45 (20 días) se obtiene un error máximo de 9.91\% (promedio de 3 corridas, básicamente da el mismo valor)

\begin{figure}[ht]
\centerline{\includegraphics[width=9cm,height=6cm]{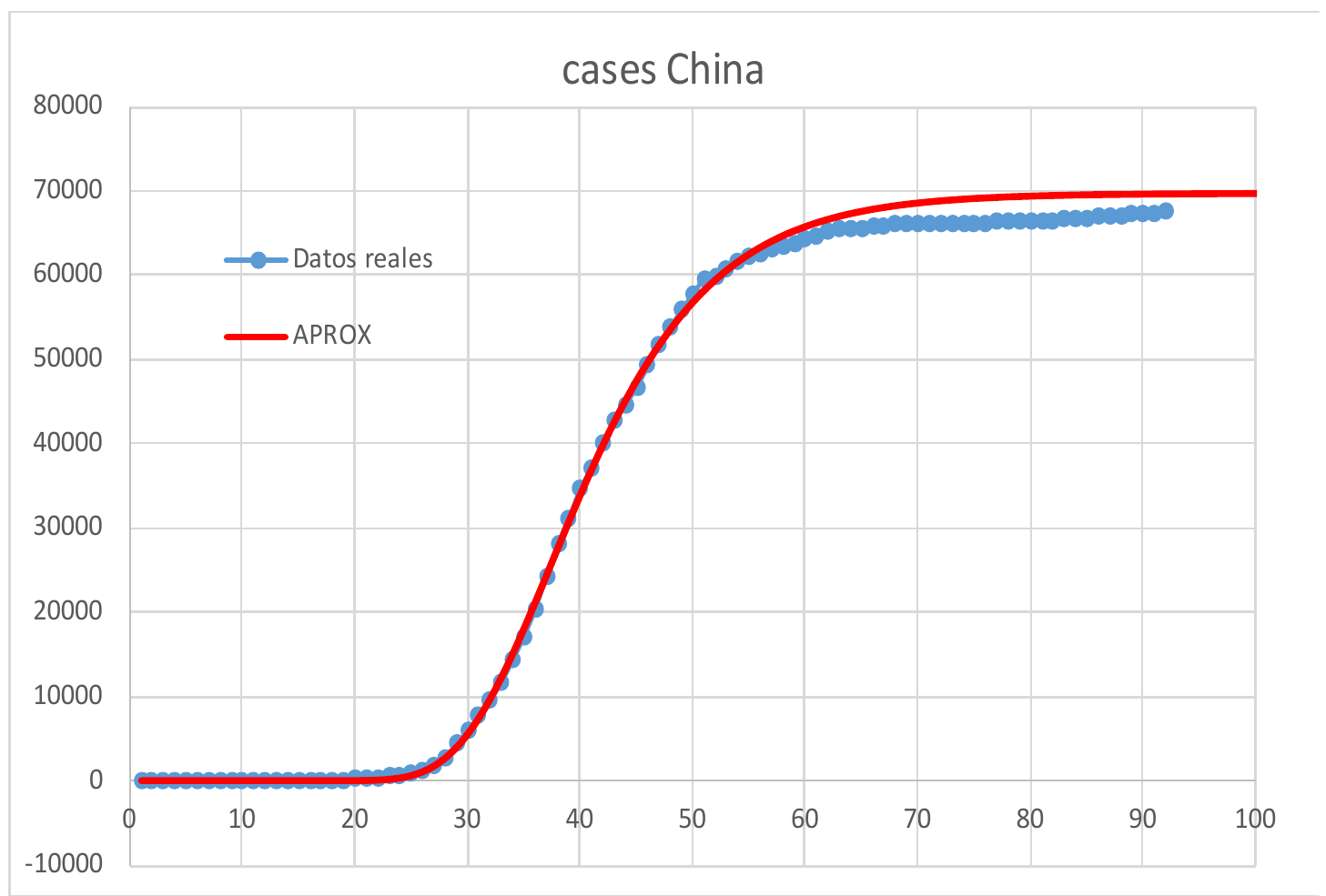}}
\caption{Predicción con Gompertz, China, 30 días }
\label{predChinaG}
\end{figure}

Al utilizar los dias de 15 al 55 (30 días) se obtiene el gráfico de la figura \ref{predChinaG}, al medir los errores relativos se obtiene un valor máximo de 4.49\%(promedio de 3 corridas), en la predicción de los datos hasta el 31/03/2020.

\begin{center}
{\sc Porcentaje de error relativo predicción  China\\ usando Gompertz}

\begin{tabular}{cc}
número de días &  máx error rel.\\\hline
20 días	&		9.91\% \\	
25 días	&		4.49\%	\\\hline
\end{tabular}
\end{center}

\subsection*{Datos de Corea del Sur}

\begin{figure}[ht]
\centerline{\includegraphics[width=9cm,height=6cm]{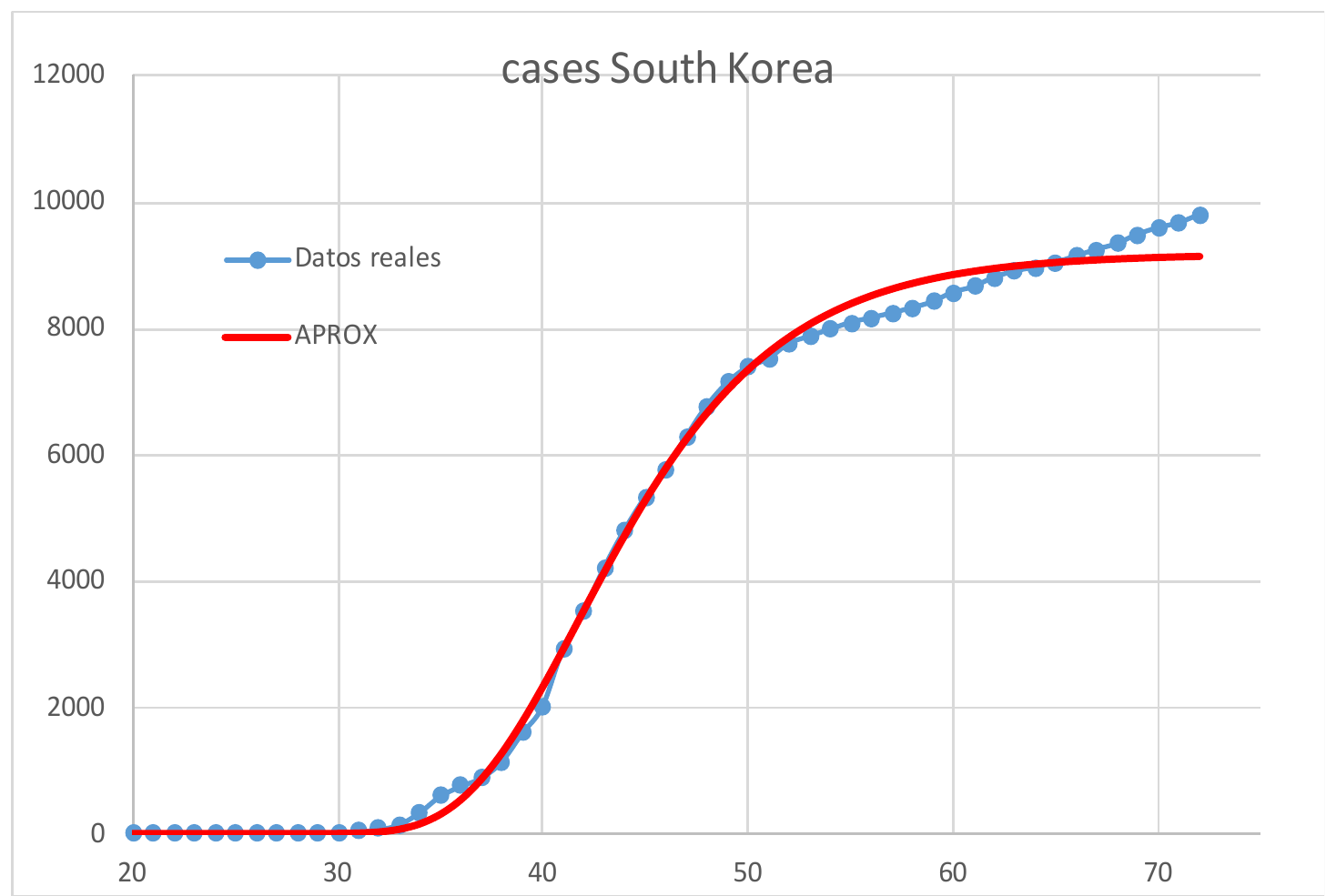}}
\caption{Predicción con Gompertz, South Korea, 25 días }
\label{predCoreaG}
\end{figure}

Para Corea del sur vamos a utilizar los datos a partir del día 28 desde el primer caso (16/02/202), este caso recuérdese que estos datos tiene una tendencia lineal en los últimos días, ver figura \ref{predCoreaG}.

\begin{center}
{\sc Porcentaje de error relativo predicción South Korea\\ usando Gompertz}

\begin{tabular}{cc}
número de días &  máx error rel.\\\hline
20 días	&		17.83\% \\	
25 días	&		6.49\%	\\\hline
\end{tabular}
\end{center}

\section{Resultados}

Como resultados de este trabajo y como hemos visto se utilizará este método para dar una predicción para los datos de algunos paises empezando por Costa Rica y luego se presentaran los datos de Italia, España, y con los ajustes que tenemos ya se tiene el de China y Corea del Sur.

\subsection{China y Corea del Sur}

Para el caso de China y Corea del Sur se puede usar los resultados obtenidos en la subsecciones \ref{ChinaO} y \ref{CoreaO}.

\newpage
\subsection{Costa Rica}

Para Costa Rica se tienen los siguientes datos a partir del día 6 de marzo en que se detectó el primer caso.
\begin{center}
{\sc Datos sobre covid-19 en Costa Rica}\bigskip

\begin{tabular}{|lcc||lcc||lcc|}\hline
&	casos &	Total & 	&	casos &	Total & 	&	casos &	Total \\
Fecha &	diarios & 	 de casos &	Fecha &	diarios & 	 de casos &	Fecha &	diarios & 	 de casos \\\hline
06/03/20 & 	1 & 	1 & 	15/03/20 & 	8 & 	35 & 	24/03/20 & 	19 & 	177 \\ 
07/03/20 & 	4 & 	5 & 	16/03/20 & 	6 & 	41 & 	25/03/20 & 	24 & 	201 \\ 
08/03/20 & 	4 & 	9 & 	17/03/20 & 	9 & 	50 & 	26/03/20 & 	30 & 	231 \\ 
09/03/20 & 	4 & 	13 & 	18/03/20 & 	19 & 	69 & 	27/03/20 & 	32 & 	263 \\ 
10/03/20 & 	4 & 	17 & 	19/03/20 & 	18 & 	87 & 	28/03/20 & 	32 & 	295 \\ 
11/03/20 & 	5 & 	22 & 	20/03/20 & 	26 & 	113 & 	29/03/20 & 	19 & 	314 \\ 
12/03/20 & 	1 & 	23 & 	21/03/20 & 	4 & 	117 & 	30/03/20 & 	16 & 	330 \\ 
13/03/20 & 	3 & 	26 & 	22/03/20 & 	17 & 	134 & 	31/03/20 & 	17 & 	347 \\ 
14/03/20 & 	1 & 	27 & 	23/03/20 & 	24 & 	158 & 	01/04/20 & 	28 & 	375 \\ 
06/03/20 & 	1 & 	1 & 	15/03/20 & 	8 & 	35 & 	24/03/20 & 	19 & 	177 \\ 
07/03/20 & 	4 & 	5 & 	16/03/20 & 	6 & 	41 & 	25/03/20 & 	24 & 	201 \\ 
08/03/20 & 	4 & 	9 & 	17/03/20 & 	9 & 	50 & 	26/03/20 & 	30 & 	231 \\ 
09/03/20 & 	4 & 	13 & 	18/03/20 & 	19 & 	69 & 	27/03/20 & 	32 & 	263 \\ 
10/03/20 & 	4 & 	17 & 	19/03/20 & 	18 & 	87 & 	28/03/20 & 	32 & 	295 \\ 
11/03/20 & 	5 & 	22 & 	20/03/20 & 	26 & 	113 & 	29/03/20 & 	19 & 	314 \\ 
12/03/20 & 	1 & 	23 & 	21/03/20 & 	4 & 	117 & 	30/03/20 & 	16 & 	330 \\ 
13/03/20 & 	3 & 	26 & 	22/03/20 & 	17 & 	134 & 	31/03/20 & 	17 & 	347 \\ 
14/03/20 & 	1 & 	27 & 	23/03/20 & 	24 & 	158 & 	01/04/20 & 	28 & 	375 \\ \hline

\end{tabular}

\end{center}

Con estos datos al hacer el ajuste con la función LG se obtiene los parámetros siguientes, ya con esta cantidad de datos básicamente se tiene un sólo óptimo: 

\begin{center}
{\sc Parámetros obtenidos para Costa Rica\\ usando Logística Generalizada}

\begin{tabular}{cccc}
M & 	a & 	b & 		c \\\hline 
886 & 	-0.0780135 & 	-4.91608521 & 		1023.90189 \\ \hline
\end{tabular}
\end{center}
Este ajuste tiene un $R^2 =	0.99858343$

\begin{figure}[ht]
\centerline{\includegraphics[width=15cm,height=9cm]{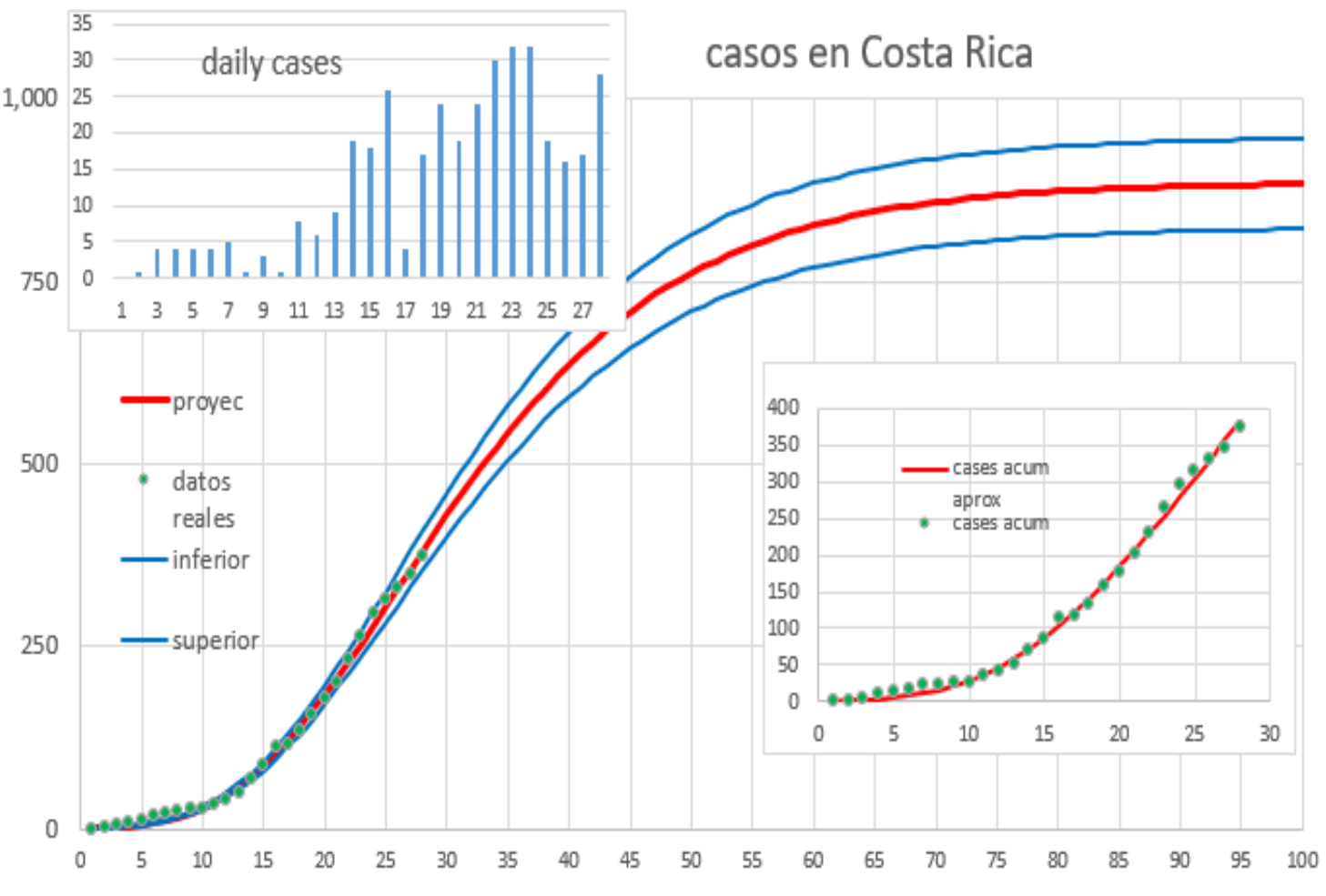}}
\caption{Predicción con LG, Costa Rica}
\label{CostaRica}
\end{figure}

En el gráfico de la figura \ref{CostaRica} se ve la predicción de casos, se puede ver el número de casos diarios en el recuadro superior y el ajuste de los datos reales en el recuadro inferior.  

Se incluyen 2 líneas de cota superior e inferior para la predicción, que se supone un 7\% de error en los datos tomado a partir de los resultados de la sección \ref{PbaPred}.

Se ve además que de seguir las situaciones como están al día de hoy habr un tope de aproximadamente 1000 casos, esto es desde el punto de vista matemático o ajuste de curvas.\bigskip
 
{\bf Gompertz}\bigskip

\begin{figure}[ht]
\centerline{\includegraphics[width=15cm,height=9cm]{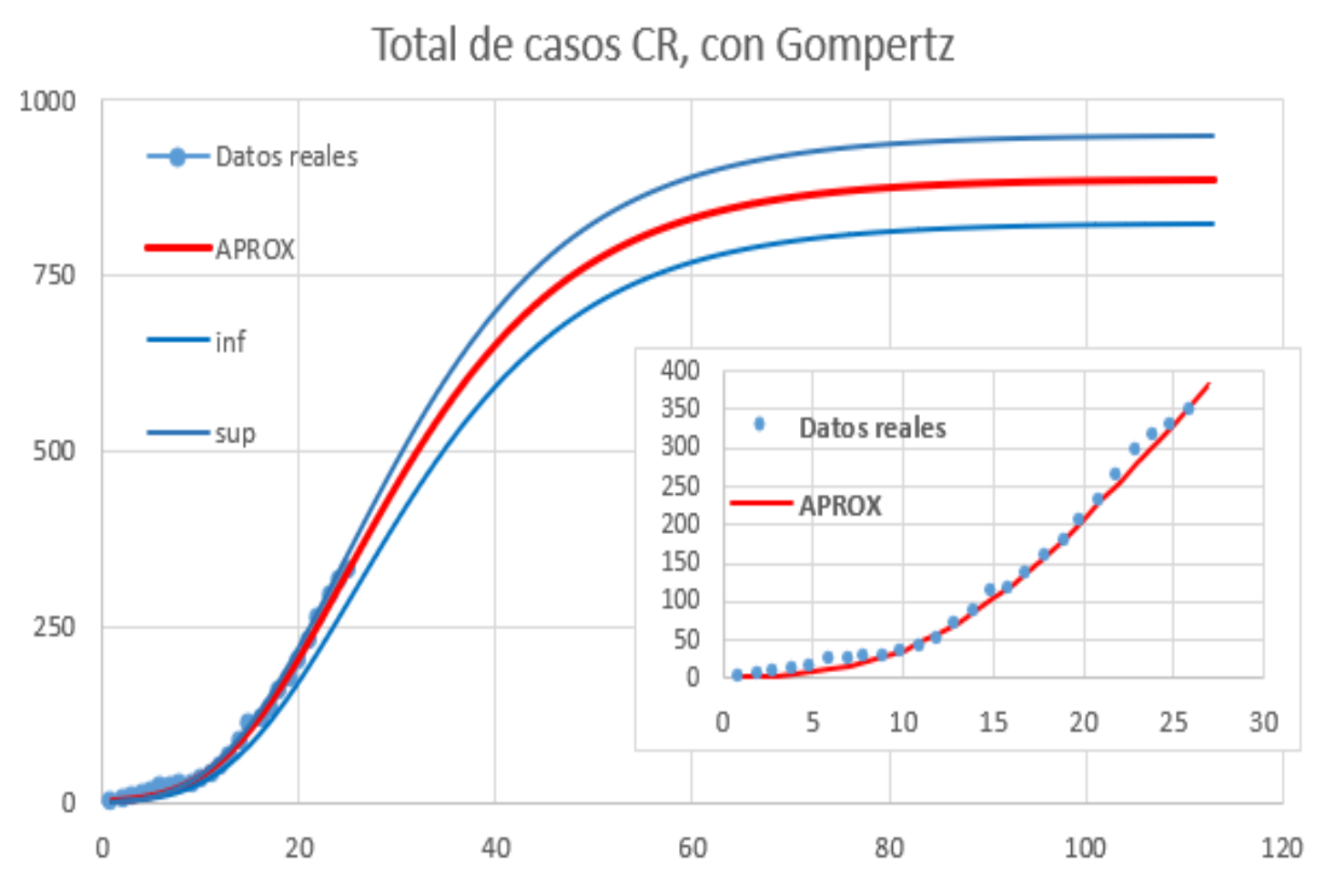}}
\caption{Predicción con Gompertz, Costa Rica}
\label{CostaRicaG}
\end{figure}

Con la curva de Gompertz se obtiene básicamente el mismo resultado como se ve a en lo que sigue:

\begin{center}
{\sc Parámetros obtenidos para Costa Rica\\ usando Gompertz}

\begin{tabular}{cccc}
 	a & 	b & 		c \\\hline 
887 & 	-6.923348908 & 		-0.077891632 \\ 
\end{tabular}
\end{center}
Este ajuste tiene un $R^2 =	0.99853504$,  en la figura \ref{CostaRicaG} se puede ver el gráfico resultante. 

Debe notarse que los resultados de utilizar estas 2 funciones se han ido aproximando entre ellas conforme pasan los días y para muestra se presenta el valor máximo entre ellas en la siguiente tabla.

\newpage
\begin{center}
{\sc Valor máximo alcanzado por las curvas \\ según día  para Costa Rica}
\\
\begin{tabular}{cccc}
	Fecha &	Gompertz &	LG \\\hline
23/03/20 & 	2162 & 	1983 \\ 
24/03/20 & 	1445 & 	1359 \\ 
25/03/20 & 	1310 & 	1274 \\ 
26/03/20 & 	1583 & 	1505 \\ 
27/03/20 & 	2048 & 	1974 \\ 
28/03/20 & 	2426 & 	2343 \\ 
29/03/20 & 	1743 & 	1730 \\ 
30/03/20 & 	1193 & 	1185 \\ 
31/03/20 & 	924 & 	922 \\ 
01/04/20 & 	887 & 	886 \\ \hline
\end{tabular}
\end{center}

\newpage

\subsection{Italia}

Para Italia se presenta los resultados por Gompertz y con los datos al 30/3/2020.

\begin{figure}[ht]
\centerline{\includegraphics[width=15cm,height=9cm]{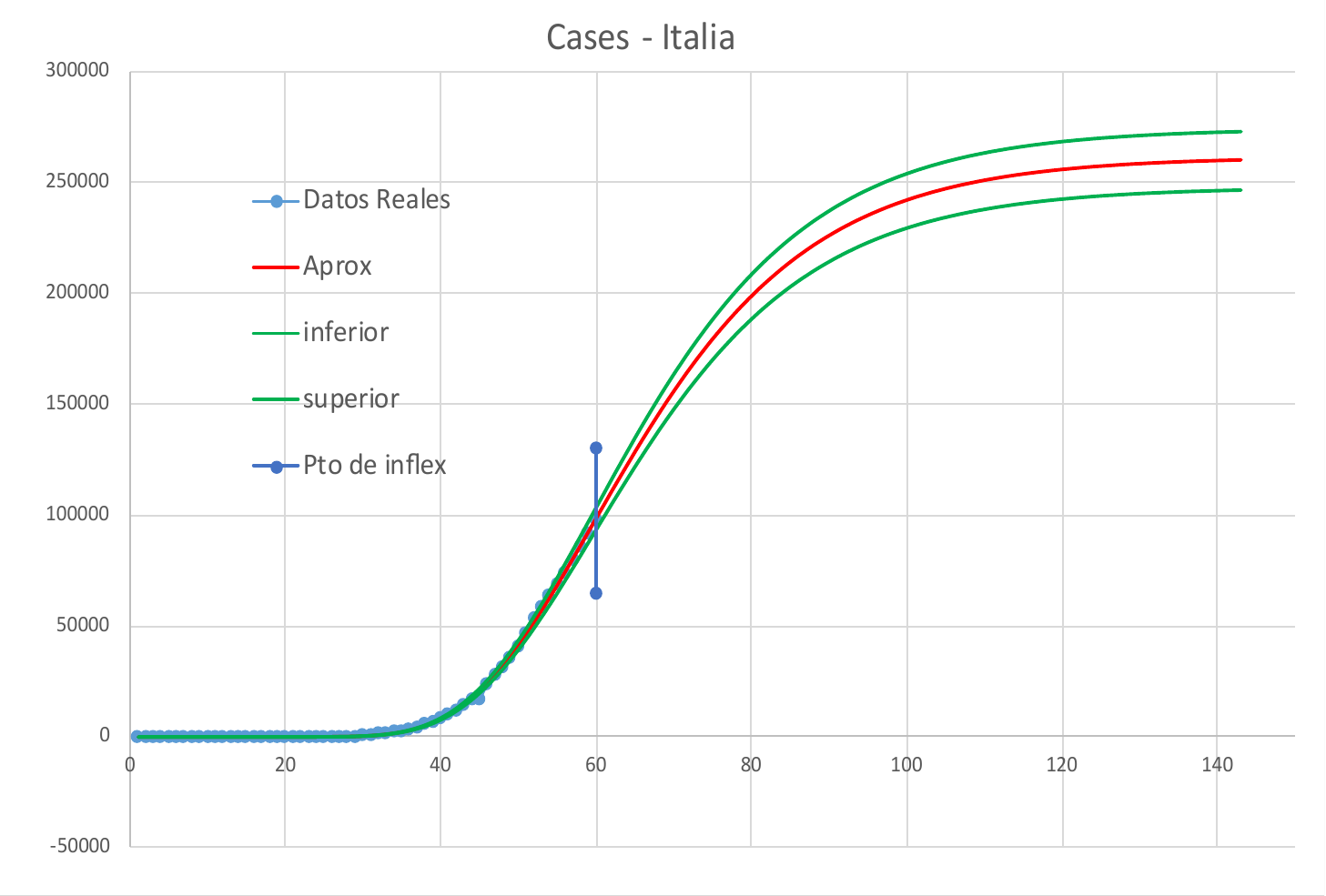}}
\caption{Predicción con Gompertz, Italia}
\label{Italia}
\end{figure}

\begin{center}
{\sc Parámetros obtenidos para Italia\\ usando Gompertz}
 
\begin{tabular}{cccc}
 	a & 	b & 		c \\\hline 
261 052 & 	-43.77482253 & 		-0.063450536 \\ 
\end{tabular}
\end{center}
Este ajuste tiene un $R^2 =	0.99968$,  en la figura \ref{Italia} se puede ver el gráfico resultante. 

\newpage

\subsection{España}

Para España se presenta los resultados por Gompertz y con los datos al 31/3/2020.

\begin{figure}[ht]
\centerline{\includegraphics[width=15cm,height=9cm]{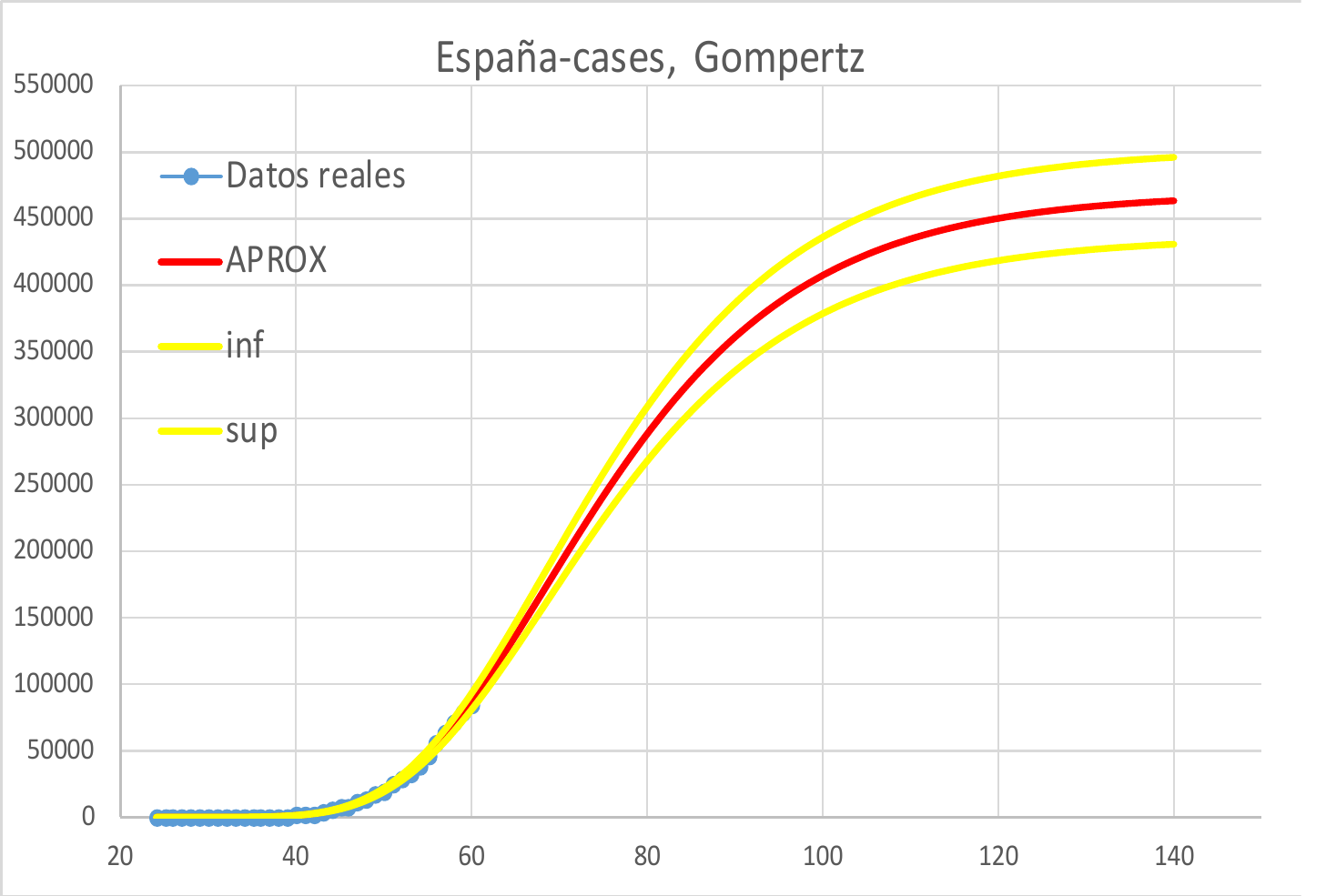}}
\caption{Predicción con Gompertz, España}
\label{Espana}
\end{figure}

\begin{center}
{\sc Parámetros obtenidos para España\\ usando Gompertz}
 
\begin{tabular}{cccc}
 	a & 	b & 		c \\\hline 
468 495 & 	-69.20043418 & 		-0.061995389 \\ 
\end{tabular}
\end{center}
Este ajuste tiene un $R^2 =	0.999410067$,  en la figura \ref{Espana} se puede ver el gráfico resultante. 

\subsection{Otros paises}

Se tienen los resultados de otros paises que se van a ir incorporando más adelante

\section{Conclusiones y trabajo futuro} 

Con este trabajo se muestra como se puede ajustar curvas de crecimiento de poblaciones, utilizando las funciones LG y Gompertz, inclusive en el caso más general como en los datos de Corea del Sur, 
sección \ref{CoreaO}, donde se está haciendo de la logística con recta.

Se ve que estos métodos podrían ser utilizados para predecir el crecimiento de poblaciones, en este caso de personas infectadas por el Covid--19, y podrían ayudar a los expertos en pandémias para tomar las medidas necesarias.

\subsection{Trabajo futuro}

Se necesita hacer más estudios para afinar los resultados de este trabajo

Ver la posibilidad de usar curvas semejantes para la aproximación de otro tipo de datos.

\end{document}